\documentclass[sigconf,natbib=false]{acmart}
\AtBeginDocument{%
  }

\setcopyright{acmcopyright}
\copyrightyear{2023}
\acmYear{2023}
\setcopyright{acmlicensed}\acmConference[CF '23]{20th ACM International Conference on Computing Frontiers}{May 9--11, 2023}{Bologna, Italy}
\acmBooktitle{20th ACM International Conference on Computing Frontiers (CF '23), May 9--11, 2023, Bologna, Italy}
\acmPrice{15.00}
\acmDOI{10.1145/3587135.3592169}
\acmISBN{979-8-4007-0140-5/23/05}

\RequirePackage[
  datamodel=acmdatamodel,
  style=acmnumeric,
  sortcites=true,
]{biblatex}

\begin{document}

\title[Towards Fast and Scalable Private Inference]{Towards Fast and Scalable Private Inference}
% \subtitle{\vspace{-.4em}
% Invited Paper
% \vspace{-.3em}}
\subtitle{Invited Paper\vspace{-.3em}}
% \authornote{Invited Paper}

% 

% \title{Characterizing and Optimizing End-to-End\\ Systems for Private Inference}
% \author{Jianqiao Mo, Karthik Garimella, Negar Neda, Austin Ebel, Brandon Reagen\\
% New York University\\
% {\{jm8782, kg2383, nn2231, abe5240, bjr5\}@nyu.edu}}
% \date{}

\author{Jianqiao Mo, Karthik Garimella, Negar Neda, Austin Ebel, Brandon Reagen}
% \email{
% {jm8782, kg2383, nn2231, abe5240, bjr5}@nyu.edu
% }
\affiliation{
    \institution{\{jm8782, kg2383, nn2231, abe5240, bjr5\}@nyu.edu \\
    New York University}
    \city{New York}
    \state{New York}
    \country{USA}
}
\date{}

%\maketitle
% \affiliation{%

% \affiliation{%

% \affiliation{%

% \renewcommand{\shortauthors}{}
\renewcommand{\shortauthors}{Jianqiao Mo, Karthik Garimella, Negar Neda, Austin Ebel and Brandon Reagen}

\begin{abstract}
Privacy and security have rapidly emerged as first order design constraints. 
Users now demand more protection over who can see their data (confidentiality) as well as how it is used (control). 
Here, existing cryptographic techniques for security fall short: they secure data when stored or communicated but must decrypt it for computation. 
Fortunately, a new paradigm of computing exists, which we refer to as privacy-preserving computation (PPC). 
Emerging PPC technologies can be leveraged for secure outsourced computation or to enable two parties to compute without revealing either users' secret data. 
Despite their phenomenal potential to revolutionize user protection in the digital age, the realization has been limited due to exorbitant computational, communication, and storage overheads.

This paper reviews recent efforts on addressing various PPC overheads using private inference (PI) in neural network as a motivating application.
% This paper reviews recent efforts on addressing various PPC overheads using private (neural) inference (PI) as a motivating application. 
First, the problem and various technologies, including homomorphic encryption (HE), secret sharing (SS), garbled circuits (GCs), and oblivious transfer (OT), are introduced. 
Next, a characterization of their overheads when used to implement PI is covered. 
The characterization motivates the need for both GCs and HE accelerators. 
Then two solutions are presented: HAAC for accelerating GCs and RPU for accelerating HE. 
To conclude, results and effects are shown with a discussion on what future work is needed to overcome the remaining overheads of PI.

\end{abstract}



%\received{28 February 2023}
%\received[accepted]{24 March 2023}

\maketitle

% Communication is secure.
% Storage is secure.
% Computation is not secure.
% We want to close the loop.

\section{INTRODUCTION}
\label{sec:Introduction}
\begin{figure}[t]
\begin{center}
\centerline{\includegraphics[width=.89\columnwidth]{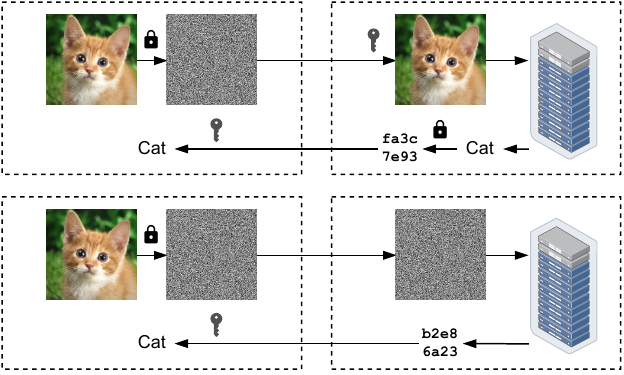}}
\vspace{-1.3em}
\caption{Upper: Standard cloud-service protects communication, but client's privacy is not preserved toward the server.
Lower: PPC enables entire privacy protection. 
Data remains encrypted during cloud computation.}
\label{fig:PPCsystem}
\vspace{-3em}
\end{center}
\end{figure}

% \fixme{1 page}
% \subsection{Motivation - Why?}
As privacy concerns continue to rise, users are demanding more protections for their data, including confidentiality and control over how it is used.
%Ideally, users could leverage cloud services without revealing their sensitive, private data to the service provider.
Current security techniques, which only protect communication and storage, do not provide sufficient guarantees for users during computation.
In other words, existing cryptographic techniques secure data when stored or communicated but must decrypt it for computation, as Figure~\ref{fig:PPCsystem} shows. 
To address this issue, PPC is a solution that can extend security guarantees to the entire execution life cycle.
% By using PPC, users can benefit from improved privacy and security while still being able to use online services integral to everyday life.
By using PPC, users can access online services with improved security guarantees for their own data.
% The challenge is that PPC incurs extremely-high performance overheads compared to plaintext execution.
However, the challenge of PPC is that it incurs extremely-high performance overheads compared to plaintext execution.
This performance gap is significant.
Protected storage and secure communication can be efficient and cheap, but not PPC.
Our goal is to overcome these overheads and realize PPCs as a practical computational paradigm with system-level optimizations and novel hardware accelerators.
%PPC requires cryptographic tools requiring high computation and communication cost, which are still open to be mitigated by novel hardware accelerators and system-level optimizations.

Deep learning is a natural starting point for demonstrating and benchmarking PPC.
It relies heavily on the client-cloud model and requires access to user data.
Additionally, neural networks account for a significant amount of private user data processing, with some companies (e.g., Meta) processing trillions of inferences per day~\cite{FacebookTrillions}.
% Ensuring the privacy of essential neural network functions, such as convolutions, ReLU, and fully-connected layers, would safeguard a disproportionate amount of users' data and mark a significant achievement in privacy-preserving computation.
Ensuring the privacy of essential neural network functions, such as convolution and ReLU, would safeguard a disproportionate amount of users' data.
PPC is still a developing field, and as we will show inference is still a ways off from being practical.
In this paper, we focus on private inference (PI) and leave training to future work.

% \subsection{Open Need}
At a high level, PI involves encrypting a user's input and (optionally, depending on the threat model) the server's neural network.
The encrypted data and model are processed using PPC methods to execute an inference on the protected user data.
%followed by the privacy-preserving computation methods over encrypted data to perform inference on the user's input.
At the end of the execution, the intended party(-ies) learns the inference result,
and neither party learns anything else about the other's input.
There exists a large body of research on PI using both homomorphic encryption and secure multi-party computation (MPC), including secret sharing and garbled circuits.
Specifically, most protocols use HE and SS for linear functions (convolutions and fully-connected layers), and GCs for nonlinear functions, including ReLUs~\cite{Sphynx, SNL, GAZELLE, DeepReDuce, delphi}.
% Non-hybrid approaches that only using HE exist, but typically sacrifice accuracy~\cite{Cryptonets, Craterlake}.
Non-hybrid approaches have been proposed.
However, they often sacrifice accuracy due to approximating activation functions with polynomials and are not considered here~\cite{Cryptonets, Craterlake, sisyphus}.

% \subsection{What we are doing}
Recent studies have focused on enhancing hybrid protocols through improved neural architecture design~\cite{Sphynx, DeepReDuce, CryptoNAS, SNL}, protocol optimization~\cite{SNL, SAFENet, GAZELLE, delphi}, and hardware acceleration~\cite{Cheetah, Craterlake, BTS, ARK, Maxelerator}. 
% These optimizations mainly aim to reduce specific components of the overall PI protocol overhead, and it is not entirely clear how they synergize to enhance the ultimate goal of fast, end-to-end private inference.
% In this paper, we start from an end-to-end perspective
State-of-the-art PI protocols break the entire PI process into two phases, a pre-processing (or offline) phase and an online phase, to move expensive computations offline and improve online inference latency.
Although there are various optimizations aiming to solve specific aspects of the hybrid PI protocols, it is not yet clear how they combine 
to enhance the end goal of fast, end-to-end PI.
% In this paper, we review recent efforts on addressing PPC overheads using PI as a motivating application.

To understand the problem, and where we as a community stand,
this paper reviews characterizations of both the online and offline phases, accounting for compute, storage, and communication overheads of the system.
First, we consider inference request arrival rates 
% (workloads of inferences) 
in PI, while most prior work focuses on individual inferences in isolation only. 
Results show it a vital distinction for computation latency, as the offline costs are so large that they do not always remain offline, even at low arrival rates.
The offline phase involves HE and GCs computations that typically require minutes to complete and it is hard to hide this latency. 
As for storage, it also incurs significant offline pressure, which can amount to tens of GBs per inference on the client.
The storage requirement constrains the number of PI pre-computations that can be buffered and can quickly exceed the storage capacities of limited client devices.
Meanwhile, ReLU activation introduces both significant computation and communication overheads for the nonlinear function evaluation.

% \subsection{Implications of our work}
We then give the insights and introduce our solutions to solve these challenges in Section~\ref{sec:solution}. 
At the system level, to overcome the limits of client storage we present the \textit{Client-Garbler} protocol.
Reversing the client and cloud roles in GCs enables pre-computed GCs to be stored on the server instead of a client's smartphone.
To address HE's computation overhead, \textit{Layer-parallel HE} (LPHE) provides embarrassing parallelism across neural network layers to significantly reduce the offline cost.
LPHE takes advantage of the fact that each neural network layer's offline HE computation is independent and can be run completely in parallel.
Finally, with the \textit{Wireless Slot Allocation} (WSA), we can optimize the default provisioning of wireless bandwidth between upload and download for PI.
The combination of the proposed optimizations improves the mean inference latency by 
1.8$\times$ over the state of the art.~\cite{hybrid_PI}
Besides the system-level optimizations, we also review two hardware accelerators, \textit{HAAC}~\cite{HAAC} and \textit{RPU}~\cite{RPU}, to accelerate GCs and HE, respectively. 
% to reduce the computation latency for the garbled circuits generation/evaluation, and offline HE computation, respectively. 
HAAC aims at reducing the computation latency for the garbled circuits generation/evaluation, while RPU targets the offline HE computation overhead.
Our result shows that HAAC achieves an average speedup of 589$\times$ with DDR4 (2627$\times$ with HBM2), and RPU provides a speedup of 1485$\times$ over a CPU.

The organization of the rest paper is as follows. 
We discuss the background of cryptographic primitives and private inference in Section~\ref{sec:Background},
% and related work can be found in Section 2 and Section 3, respectively. 
characterize the problem in Section~\ref{sec:Problem}.
In Section~\ref{sec:solution}, we introduce performance solutions from three perspectives.
Related work is reviewed in Section~\ref{sec:Related}. 
Finally, we discuss our thoughts on the state and future of systems for PI in Section~\ref{sec:Conclusion}.
% give the experiment results. Section 6 concludes the paper.

\section{BACKGROUND}
\label{sec:Background}
% \fixme{1 Page, Cambridge, Karthik, Austin, Negar}

A common convolutional neural network (CNN) contains two types of operations: linear (convolutional or fully-connected layers) and nonlinear (e.g., ReLU).
% In this section, we take the common convolutional neural network (CNN) as an example to introduce its component and explain how to compute them privately. 
In this section, we introduce the basics of how to compute the linear and nonlinear functions privately as well as complete private inferences.

% \subsection{Neural Network Inference}

% A CNN takes an input and processes it through a sequence of linear and non-linear layers in order to classify it into one of the potential classes. 

% \textit{Linear Layers}:
% The linear layers contains two types: convolutional layers or fully-connected layers.

% \textit{Nonlinear Layers}:

\subsection{Private Linear Computation (HE, SS)}
Two common linear computations in neural network inference are fully-connected and convolution layers, both of which can be expressed as matrix-vector multiplication.
Here, the matrix represents the trainable parameters of the linear layer and the vector represents the input data. 
Given that linear transformations can be efficiently expressed as arithmetic circuits with bounded depth, a natural cryptographic primitive for privately computing the linear layers on encrypted inputs is HE.

HE is an encryption scheme that allows for computation directly on encrypted data without the need for decryption, thus preserving data confidentiality during the actual computation. 
% More specifically, leveled homomorphic encryption (LHE) schemes support arithmetic operations such as addition and multiplication in ciphertext space and are thus suitable for performing encrypted linear layer evaluations. 
Given two parties, a client (with some data) and a server (with a linear function), the client can first encrypt their data under a HE scheme and send their encrypted data to the server.
The server can then homomorphically evaluate the linear function on the client’s ciphertext without ever decrypting the client's data server-side. 
The resulting linear computation produces a ciphertext that can be sent back to the client who can decrypt the ciphertext locally.  

HE introduces a large computational overhead of 4-6 orders of magnitude slowdown over their plaintext counterparts (1080 seconds for a single ResNet-18 inference \cite{hybrid_PI, Craterlake}).
Rather than directly using HE for computing linear layers, it is common to combine HE with Additive Secret Sharing (SS), another cryptographic building block that supports plaintext-level speeds for computing linear layers. 
In this HE+SS setting, linear layers are processed in two steps: a pre-processing or offline phase independent of the client's input and an online phase dependent on the client's input. 
In the pre-processing phase, the server performs a homomorphic evaluation of the linear layer on a randomly sampled and encrypted input to generate the client's additive secret share. 
This pre-processing step enables the server to perform a simple secret share evaluation of the linear layer on the client's actual data during the online phase.

\subsection{Private Nonlinear Computation (GCs)}
% The non-linear layers consist of an activation function that acts on each element of the input separately or a pooling function that reduces the output size. 
% Typical non-linear functions can be one of several types: the most common in the convolutional setting are max-pooling functions and ReLU functions.
% Nonlinear layers are a crucial component in deep learning models, as they introduce nonlinear transformations to the input data. 
% These layers commonly include either an activation function, which applies a mathematical operation to each element of the input individually, or a pooling function, which reduces the size of the output. 
Nonlinear layers are a crucial component in deep learning models, which include either an activation function or a pooling function. 
The activation function is typically chosen from several types of nonlinear functions, such as the widely-used rectified linear unit (ReLU) function. 
In convolutional neural networks, max-pooling functions are also frequently employed.

% The key observation that we wish to make in this context is that all these functions can be implemented by circuits that have size linear in the input size and thus, evaluating them using conventional 2PC approaches does not impose any additional asymptotic communication penalty.
Generally, HE and additive SS enable private arithmetic computations, i.e., additions and multiplications over integer values.
Garbled Circuits~\cite{Yao1986GC} enable two parties to compute a Boolean function on their private inputs without revealing their inputs to each other.
Unlike SS and HE, operating over Boolean gates enables the parties to jointly compute \textit{arbitrary} function, making GCs a solution to privately compute nonlinear layers. 
We note that binary constructions of HE exist (e.g., TFHE~\cite{2020TFHE}).
However, these incur extremely high performance overheads with prior work. 
% \fixme{reporting 75-600ms per Boolean AND gate{cites}.}
Evaluating single Boolean gate with HE can take 75-600ms to process~\cite{Stochastic_TFHE, FHEW_like}.

In GCs, a function is represented as a Boolean circuit.
% One party (the garbler) assigns random labels to each input wire of each gate.
% For each gate, the garbler then generates an encrypted truth table that maps the output labels to the gate’s input labels. 
% The garbler sends the generated garbled circuit to the other party (the evaluator), along with the labels corresponding to its inputs. 
One party (the garbler) assigns random labels representing \{0, 1\} to each input wire of each gate, generates an encrypted truth table that maps the output labels to the gate's input labels.
% The evaluator then uses the OT protocol~\cite{ObliviousTransfer} to obtain the labels corresponding to its inputs without the garbler learning the input values. 
The evaluator uses Oblivious Transfer~\cite{ObliviousTransfer} to obtain labels corresponding to its inputs without revealing the values to the garbler, and then executes the circuit gate-by-gate.
% At this point, the evaluator can execute the circuit in a gate-by-gate fashion, without learning intermediate values. 
Finally, the evaluator shares the output labels with the garbler who maps them to plaintext values.
Recent algorithmic optimizations are used to construct high-performance GCs \cite{FreeXOR,Half_Gate}.
More details can be found in~\cite{Half_Gate_rekey, GCintro2, GCintro1}.

\subsection{Private Inference}
\label{subsec:pi_delphi}
\textit{Starting point}.
We briefly describe DELPHI~\cite{delphi}, a baseline hybrid PI protocol that uses HE and SS for linear layers, and GCs for ReLU layers.
In the 2-party-computation setting, a client uses her data for inference on  the server’s proprietary model, without learning the server’s model parameters.
The DELPHI protocol consists of an offline phase, which only depends on the network architecture and parameters, and an online phase, which is performed after the client’s input is available.
In this way, many expensive computations are moved to offline to improve online inference latency.

\textit{Offline Phase}.
First, the client sends their public keys to the server.
At the $i$th layer, client and server sample random vectors $r_i$ and $s_i$ respectively.
Then the client sends the encrypted random vector $E(r_i)$ to the server.
The server uses its model parameter $w_i$ to compute $E(w_i\cdot r_i-s_i)$ homomorphically and returns to the client, thus the client holds the share $ \langle y_i \rangle _c=w_i\cdot r_i-s_i$.
% a  encryption of its share of the linear transformation and sends it back to the client.
% The client and the server then sample random vectors per each linear layer from a finite ring. 
% The client encrypts its random vectors and sends them to the server. 
% The server then computes a homomorphic encryption of its share of the linear transformation and sends it back to the client.
The server also constructs GCs for each nonlinear ReLU operation 
% that implements $ReLU(y_i)-r_{i+1}$, 
and sent them to the client. 
The client obtains labels corresponding to its $ \langle y_i \rangle _c$ and $r_{i+1}$ using OT, where $r_{i+1}$ is prepared for the next $(i+1)$th layer.
% Here $y_i$ is the output of linear operation 
% The server also constructs and garbles a circuit for each ReLU operation, which is sent to the client, who obtains labels for its inputs using an OT protocol. 
% The communication and storage requirements are also shown for ResNet-18 inference on a single input from TinyImageNet.

\textit{Online Phase}.
After the client’s input $x_1$ is available, the client sends the subtraction $x_1-r_1$ to the server.
At the beginning of the $i$th layer, the server uses the subtraction to calculate its share of the layer output $ \langle y_i \rangle _s=w_i(x_i-r_i)+s_i$. 
To evaluate the ReLU layer, the server (garbler) sends the labels corresponding to its $ \langle y_i \rangle _s$ to the client.
The client uses its labels of $ \langle y_i \rangle _c$ and $r_{i+1}$, together with the label of $ \langle y_i \rangle _s$ to evaluate the garbled circuits ReLU$(\langle y_i \rangle _c + \langle y_i \rangle _s)-r_{i+1}$. 
The GCs' output labels will be sent to the server as they represent $(x_{i+1}-r_{i+1})$.
At this point, the client holds the share $r_{i+1}$ and the server holds $(x_{i+1}-r_{i+1})$. 
They will similarly evaluate subsequent layers.

% the client inputs the data and computes the subtraction of a random vector from it. 
% This subtracted value is sent to the server, who uses it to calculate its share of the layer output using a linear transformation. 
% The client holds the labels for the server's input, as well as the labels for its inputs, and evaluates the circuit using an OT protocol. 
% The output labels are then sent back to the server who can obtain the value of the next layer by subtracting the corresponding random vector from it.
\section{THE PI PROBLEM}
\label{sec:Problem}

% \fixme{Is figure 1 supposed to be referenced here?}

We identify three system-level bottlenecks that hinder private inference protocols from being both scaled and deployed: high client-side storage costs, client- and server-side compute latency, and communication costs.

\textbf{Storage}: We observe that ReLU GCs are the primary storage bottleneck in PI.
The common setup is to generate garbled ReLUs server-side and store them on the client device during the offline phase. 
We benchmark the fancy-garbling library~\cite{swanky_fancyGC} to understand the costs. 
For each scalar ReLU operation, the garbler (the server) must store 3.5KB per ReLU while the evaluator (the client) pays a penalty of 18.2KB per ReLU. To put this in perspective, to perform \emph{a single inference} using a high-performance network (ResNet-18) on TinyImageNet requires the client to store 41GB of garbled ReLUs. For the same network on a larger dataset (ImageNet), the client must store 498GB of ReLU GCs. 

\textbf{Compute}: In the baseline protocol described above, the server performs ReLU GCs garbling and both HE and SS evaluation of linear layers. Meanwhile, the client is responsible for ReLU GCs evaluation.  
Secret shares generated via HE as well as GCs garbling are performed in the offline phase, leaving only GCs evaluation and the server’s SS evaluation as the computation that must happen online. 
Our analysis reveals that the server-side HE evaluations dominate the compute time of PI protocols. 
Furthermore, both GCs garbling and evaluation have non-negligible compute latency.
For example on ResNet-18 on TinyImageNet, the HE portion of these PI protocols takes 1080 seconds while the GCs portion takes 225 seconds in total. 
Crucially, we find that the GCs garbling and evaluation compute latency depends significantly upon the device performing the computation. For the above protocol, GCs garbling takes only 25 seconds on the server while GCs evaluation takes 200 seconds (when using an Intel Atom board as a client and an AMD 32-core machine as the server). 

\textbf{Communication}: Hybrid PI protocols require several rounds of communication both in the offline and online phase with the number of rounds growing with network depth. 
The transmission of ReLU GCs from the garbler to the evaluator during the offline phase dominates the communication latency in hybrid PI.
Transmitting the 41GB of garbled ReLUs needed for ResNet-18 on TinyImageNet at 5G (1Gpbs) takes roughly 747 seconds. Figure~\ref{fig:performance} (Baseline) shows the total end-to-end latency of performing a single private inference on ResNet-18 with a TinyImageNet input.

% \begin{figure}[h!]
% \begin{center}
% \centerline{\includegraphics[width=\columnwidth]{figures/standard_protocol.pdf}}
% \vspace{-1em}
% \caption{Delphi on TinyImageNet ResNet18}
% \label{fig:delphi}
% \vspace{-1em}
% \end{center}
% \end{figure}

\section{OUR SOLUTIONS}
\label{sec:solution}

\begin{figure*}[t]
\centerline{\includegraphics[width=.97\textwidth]{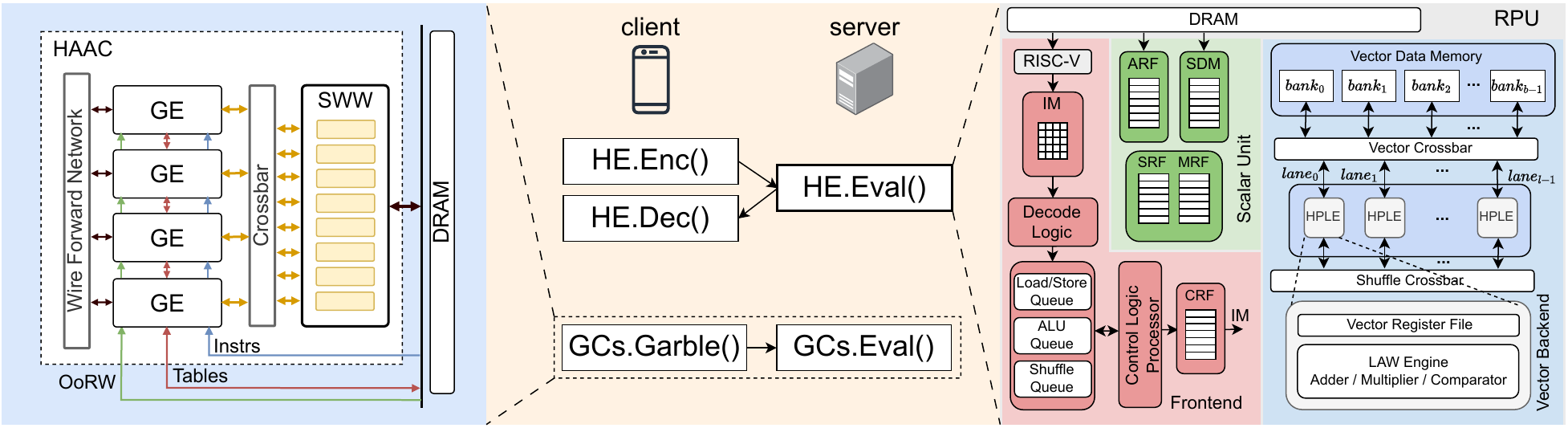}}
\vspace{-1.5em}
\caption{
Our solutions to faster PI. 
In the center, we show the main steps in PI, which include the \textit{Client-Garbler} and \textit{LPHE} as system-level optimizations. 
HE computations can be accelerated by \textit{RPU} on the right.
Both the GCs garbling and evaluation can be accelerated by \textit{HAAC} on the left.
}
% solution hardware and system
\label{fig:solution}
\vspace{-1.5em}
\end{figure*}

This section presents our end-to-end characterization of the above protocol. In particular, we uncover and mitigate three main system-level bottlenecks that prevent PI from scaling: storage, computation, and communication.
Even after our system-level optizations, we find high compute latency that necessitates custom architecture.  
We discuss two accelerators: one for garbled circuits (HAAC) and one for homomorphic encryption (RPU). 

% \begin{figure*}[t]
% \centerline{\includegraphics[width=\textwidth]{figures/System_RPU_HAAC-system.drawio.pdf}}
% \vspace{-1em}
% \caption{
% \fixme{Our solutions to faster PI. 
% }
% % solution hardware and system
% }
% \label{fig:solution}
% \vspace{-1em}
% \end{figure*}

\subsection{System-Level Design} 
\label{sec:system-design}

\textbf{Addressing Storage}: Storing the garbled ReLUs client-side during the offline phase inhibits the client from (at best) storing more than a few precomputes especially for deeper and larger networks. To overcome this limitation, the Client-Garbler protocol switches the role of the garbler and evaluator so that the server (now the evaluator) must store the ReLU GCs while holding the exact same security guarantees. This reduces the storage cost of the client device by $5 \times$; rather than storing 41GB for TinyImageNet inference on ResNet-18, the client now only stores 8GB per inference. Outside of storage costs, Client-Garbler also reduces online phase latency as the powerful server is now evaluating the garbled ReLUs while the offline phase latency increases as the client must now perform garbling. This latency tradeoff is reflected in Figure \ref{fig:performance} (+SysOpt).

\textbf{Addressing Compute}: Server-side HE evaluations of the linear layers during the offline phase account for a majority of the compute cost. As discussed in Section \ref{subsec:pi_delphi}, the client and server iterate through each linear layer to generate secret shares that are used for fast, online phase linear evaluations. These secret shares are generated independently for each linear layer meaning each linear layer HE evaluation can be run in an embarrassingly parallel manner on the server which we call Layer-Parallel Homomorphic Evaluation (LPHE). Rather than performing each independent HE evaluation sequentially on a single core, each HE evaluation is allocated to a separate core server-side. For ResNet-18 on TinyImageNet, LPHE reduces the HE evaluation run-time from 1080s to just 141s. Across all datasets and networks, LPHE speeds up HE by $9.7 \times$.

\textbf{Addressing Communication}: The transmission of garbled ReLUs from the garbler to the evaluator in the offline phase dominates the communication latency of the entire PI protocol. Furthermore, the massive storage penalty per ReLU (18.2KB) results in an asymmetry in the amount of data sent between the two parties. For example, in the Client-Garbler protocol, 83.5\% of the total amount of data transferring is caused by uploading data to the server. Current 5G wireless standards allows for flexibility in the amount of bandwidth allocated to both uplink and downlink. By optimally allocating the bandwidth split (Wireless Slot Allocation), the communication latency of PI protocols can be reduces by 35\%.

Putting these aforementioned system-level optimizations helps us not only reduces the storage cost of the client but also reduce both the offline and online phase latency. Figure \ref{fig:performance} (+SysOpt) shows the latency breakdown with our optimized protocol. Compared to Figure \ref{fig:performance} (Baseline), these optimizations reduce the end-to-end latency from 2050 seconds to 1052 seconds. Even after these optimizations, high compute costs from both the HE and GCs portions of the protocol dominate the runtime and necessitate custom architecture to further reduce latency.

% \begin{figure}[H]
% \begin{center}
% \centerline{\includegraphics[width=\columnwidth]{figures/optimized_protocol.pdf}}
% \vspace{-1em}
% \caption{Optimized Protocol on TinyImageNet ResNet18}
% \label{fig:cryptonite}
% \vspace{-1em}
% \end{center}
% \end{figure} 

% \fixme{Figure~\ref{fig:performance} (b) shows the performance gain from the system-level optimization. 
% The overall latency is significantly reduced thanks to the optimized protocol.
% In the offline phase, the HE evaluation that occupied the majority of the latency in the baseline is greatly decreased, as well as the GCs evaluation for the online phase.
% However, besides communication, the computation for HE and GCs (i.e., the HE evaluation and GCs garbling) now dominate the total latency, which motivate us to look for a faster HE and GCs solution from other aspects.
% % the total end-to-end latency of performing a single private inference on ResNet18 with a TinyImageNet input. 
% % During the offline phase, the HE evaluations dominate the runtime while during the online phase, most compute cycles are spent on client-side GC evaluation. 
% }

% \begin{figure*}[t]
% \centerline{\includegraphics[width=\textwidth]{figures/System_RPU_HAAC-system.drawio.pdf}}
% \vspace{-1em}
% \caption{
% \fixme{Our solutions to faster PI. 
% In the middle we show the main steps in PI.
% The HE evaluation can be accelerated by RPU on the right.
% Both the GCs garbling and evaluation can be accelerated by HAAC on the left.
% }
% % solution hardware and system
% }
% \label{fig:solution}
% \vspace{-1em}
% \end{figure*}

\subsection{Accelerating GCs with HAAC}
\label{sec:HAAC}

Given the system-level optimization mentioned above, the GCs storage stress of the client side is largely relieved, but it raises the pressure on GCs computation.
The GCs accelerator should be not only fast 
% and lightweight, but also 
% \fixme{efficient--you mean energy?} 
but also lightweight and energy efficient.
It should be able to process large amounts of data and maintain high throughput. 
% Here we introduce a novel GC accelerator with a specialized compiler, named HAAC~\cite{HAAC}.
Here, we present HAAC~\cite{HAAC}, a novel hardware-software co-design that includes a compiler and hardware accelerator that combine to improve GCs performance and efficiency, making PI more practical.
% HAAC is a hardware-software co-design that accelerates GCs to mitigate performance overheads and make PI more practical. 
% The approach includes a compiler, ISA, and hardware accelerator that combine to significantly improve GC performance and efficiency.
% The design philosophy of HAAC is to keep hardware simple and efficient, maximizing area devoted to custom execution units and other circuits essential for high performance. 

% \fixme{--Is this text origional?--}
Employing hardware-software co-design in GCs is appropriate, as the programs including all dependence, memory accesses, and control flow, are already known and fixed at compile time.
% GCs are exemplars of co-design as the programs, including all dependence, memory accesses, and control flow, are already known and fixed at compile time.
Potential benefits in hardware can be realized through optimization on the software side. 
HAAC's design philosophy is to keep hardware simple and efficient, maximizing area utilization of custom execution units and other circuits essential for high performance.
Our approach leverages the properties of GCs to express arbitrary programs as streams, which simplifies hardware and enables complete memory-compute decoupling.
The evaluation results demonstrate the effectiveness of the proposed approach in accelerating GCs and mitigating performance overheads.
HAAC hardware comprises Gate Engines (GE) execution units, see Figure~\ref{fig:solution}.
GE is deep, in-order pipeline that provide high performance potential, which ideally computes a gate per cycle.
However, exploiting GEs parallelism while keeping hardware efficient is a challenge.
% Considering GCs properties that the gate-dependence is known at compile time, we push the parallel scheduling to the software. 
This presents a prime opportunity for hardware-software co-design.
HAAC's compiler leverages Instruction-level Parallelism (ILP) to improve intra/inter-GE parallel processing, with instructions and constants streamed to each GE using queues.
The compiler analyzes the leveled data dependence graph for the entire baseline program to expose all available ILP.
% The compiler performs the full reordering by analysing the leveled dependence graph of the entire program, in order to expose all available Instruction-level Parallelism (ILP) and maximize instruction parallelism.
% but exploiting gate parallelism and orchestrating data effectively while keeping hardware efficient is a challenge. 
% The compiler can express GCs as multiple streams, leveraging instruction-level parallelism to improve intra/inter-GE parallel processing. 
% Queues are used to stream instructions to each GE in the right order.
% The compiler can express GCs as multiple streams, leveraging Instruction-level Parallelism (ILP) to improve intra/inter-GE parallel processing. 
This strategy, called full-reordering, works well to resolve data hazards, and significantly increases parallel GEs performance, but it can reduce the on-chip data reuse and burden the off-chip traffic.
To better balance data reuse and ILP, HAAC also purposes another scheme called segment-reordering.
Rather than computing the ILP graph for the entire program, we partition the program into contiguous parts (segments) and reorder instructions within each segment.
This provides more instruction parallelism than baseline programs and generally captures more data reuse than full-reordering.

%%%%%%%%%%%%%%%%%%%%%%%%%%%%%%%%%%%%%%%%%%%%%%%%%%
% HAAC includes a gate engine (GE) that accelerates individual gate execution. 
% GEs provide high performance potential, but exploiting gate parallelism and orchestrating data (operands, constants, instructions) effectively while keeping hardware efficient is a challenge. 
% The compiler can leverage instruction-level parallelism to improve intra/inter-GE parallel processing, with instructions and constants streamed to each GE using queues.

% One key insight is how co-design enables expressing arbitrary GCs programs as streams, which simplifies hardware and enables complete memory-compute decoupling. 
% The approach develops a scratchpad that captures data reuse by tracking program execution, eliminating the need for costly hardware managed caches and tagging logic. 
%%%%%%%%%%%%%%%%%%%%%%%%%%%%%%%%%%%%%%%%%%%%%%%%%%
With the help of HAAC's distinct on-chip memory subsystem, the high parallelism in GEs can be actually achieved instead of being blocked by input/output data latency.
% Another key insight is how co-design enables complete gate execution and off-chip accesses decoupling while also capturing on-chip data reuse.
The gate inputs and outputs, called wires data, are more difficult as they do not follow a pattern.
HAAC uses a scratchpad with multiple memory banks to capture wires reuse by tracking program execution, eliminating the need for costly hardware-managed caches and tagging logic.
The on-chip scratchpad memory, called sliding wire window (SWW), stores a contiguous address range of wires, and the range increases as the program executes. We also leverage GCs input/output patterns and strides wires across SWW banks to reduce conflicts.
Renaming is a complementary compiler pass that sequentializes gate output wire addresses according to program order. 

The SWW and renaming combine to filter off-chip accesses, as recently generated wires are often soon reused when they are still on-chip, with the performance benefits of a cache and the efficiency of a scratchpad.
HAAC's co-design also enables complete gate execution and off-chip accesses decoupling.
Used wires, marked by the compiler, will not go off-chip by Eliminating Spent Wires (ESW), which elides redundant writes to off-chip memory.
The unused wires, or Out-of-range Wire (OoRW), will be stored in the off-chip memory. 
% pushed to the queue as Out-of-range Wire (OoRW) when they are used by the later instructions. 
The compiler knows which wires will be OoR and can push them on-chip to an OoRW queue, and the GEs will to check the OoRW queue if the inputs are not on-chip.
The optimizations allow the overlap of computation and off-chip accesses, hiding the latency of data movement.

We evaluate HAAC thoroughly with benchmarks from VIP-Bench~\cite{VIP-Bench}. 
With 16 GEs, a 2MB SWW on-chip memory, and DDR4, HAAC provides an average speedup of 589$\times$ than an Intel Core i7-10700K CPU (2627$\times$ with HBM2) in 4.3mm$^2$. Figure~\ref{fig:performance} (+HAAC) shows that after accelerating GCs with HAAC, we further reduce the total latency for a private inference on ResNet-18 by 39\%.
The percentage of GCs in the computation latency is already very low, leaving the HE evaluation as a large occupation.

\subsection{Accelerating HE with the RPU}
\label{sec:RPU}
We now turn to the final optimization: accelerating expensive HE operations required for linear layers in the offline phase. 
Like RELUs, simple plaintext operations see large overheads in the encrypted domain. 
Specifically, linear transformations map to a series of complex ciphertext rotations and key switching operations, which dominate the runtime of HE, requiring frequent applications of the (inverse) number theoretic transform, or (i)NTT. Fortunately, this implies high degrees of data-level parallelism, which can be efficiently exploited with the right architectural design.

With this in mind, we now discuss our Ring Processing Unit (RPU)~\cite{RPU}.
%, a vector architecture and its associated instruction set architecture (ISA) B512. 
The RPU implements B512, a vector ISA that has been specifically designed to cater to common HE operations while also remaining well-suited to general-purpose programming. 
The decision to develop a vector architecture and ISA, rather than fixed ASIC hardware, gives us the reprogrammability needed to remain viable as HE algorithms continue to develop and mature. 

%\subsubsection{Instruction Set Overview}

\begin{figure}[t]
% \begin{figure}[h!]
\begin{center}
\centerline{\includegraphics[width=.95\columnwidth]{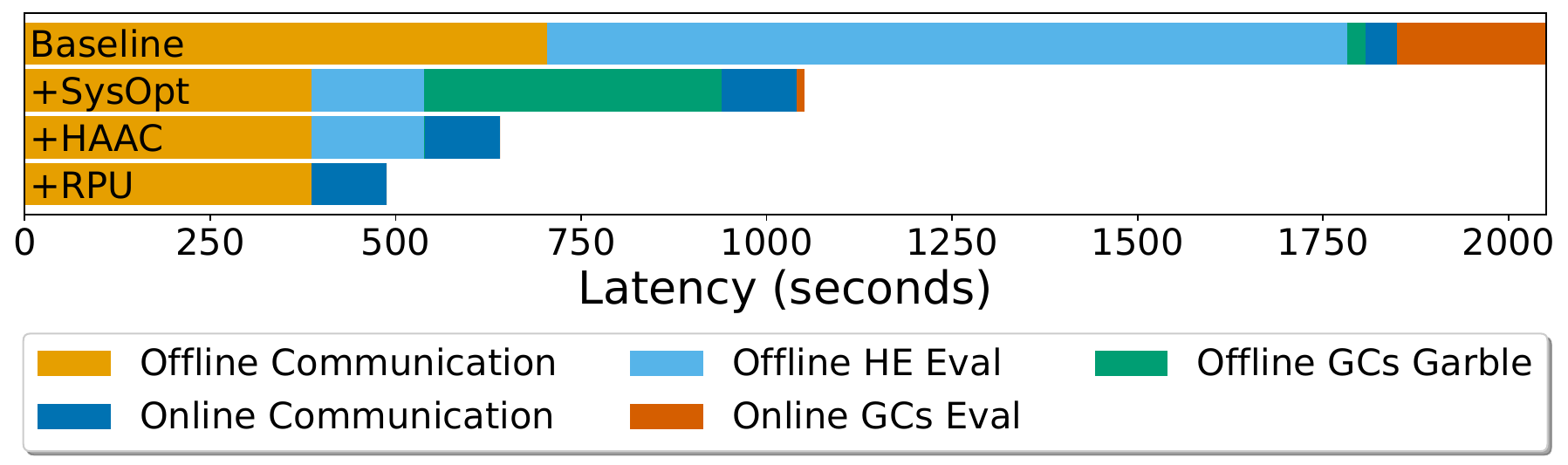}}
\vspace{-1em}
\caption{Latency of a single private inference.
The first bar shows the Baseline without optimization.
After that, system-level optimization, GCs acceleration and HE acceleration are appended successively.}
\label{fig:performance}
\vspace{-3.2em}
\end{center}
\end{figure} 

The vector architecture state includes 64 128-bit vector registers, 64 128-bit scalar registers, a 4 MiB vector data memory (VDM) expandable up to 32 MiB, and a 2 MiB scalar data memory (SDM). 
Additionally, there is an Address Register File (ARF) for indirect memory access, a Modulus Register File (MRF), and several control registers for increased flexibility. 
We operate on fixed vector lengths of 512 elements to balance microarchitectural concerns with the power-of-two input sizes common to both (i)NTT and HE kernels. 
Instructions are grouped into four categories: compute, memory, shuffle, and control.
Compute instructions perform point-wise modular computations between two vectors or a vector and a scalar value. 
Memory operations interact with the VDM and SDM to bring data to and from the vector and scalar register files respectively. 
We support four addressing modes in memory instructions to efficiently handle the complex access patterns in the (i)NTT and other HE workloads. Shuffle instructions are heavily used by (i)NTT and are efficiently supported in the microarchitecture. 
Control instructions act on the Control Register File (CRF) and drastically reduce our required code size. Overall, B512 introduces 28 instructions to meet the needs of HE-specific applications without sacrificing the generality that traditional vector operations provide.

%\subsubsection{Microarchitecture Overview}

The RPU is designed for general ring processing with high performance by taking advantage of regularity and data parallelism. 
We achieve this balance by designing explicitly managed hardware to elide the high costs and complexity of caches, dynamic scheduling logic, and prediction, and task the compiler with handling scheduling and data movement at compile time. 
Figure~\ref{fig:solution} presents the RPU microarchitecture, which consists of a front-end to handle instruction fetching, decoding, and control logic, and a backend, which provides the high-performance hardware needed to efficiently perform our HE-tailored vector operations. 
The front-end includes three decoupled queues that operate independently on compute, memory, and shuffle instructions. 
Once an instruction is in its respective queue, it can run in parallel with other instruction types, and the microarchitecture guarantees that data hazards are avoided. The parallel execution via these decoupled pipelines is key to achieving high performance with general-purpose processing as it masks much of the data movement time.

Compute instructions are passed to computational units that we denote as high-performance large arithmetic word (LAW) engines (HPLEs). 
HPLEs each operate on a slice of the 512 element vector whose size is determined by the number of parallel HPLEs (i.e., lanes) in our vector architecture. 
Each HPLE consists of a 128-bit modular multiplier, adder, subtractor, and two comparator units. 
HPLEs can be modified to support bit-serial computation, trading off performance for area.
%\fixme{The use of bit-serial computation in HE is an active line of research we intend to pursue further.}

%\subsubsection{Evaluating the RPU}

We use the SPIRAL \cite{Spiral} to map instructions onto the RPU and automatically generate high-performance B512 programs. 
SPIRAL has a rich library of transformations and optimizations that can expertly generate high-performance code across various platforms and kernels, especially in linear transforms such as the (i)NTT. 
We evaluate and characterize RPU using a detailed cycle-level simulator that is parameterized to consider a range of IP, namely modular multiplier designs, number of HPLEs, and VDM partitioning strategies. 
This enables rapid design space exploration to quantify design decisions. 
The simulator is further verified with a complete RTL implementation of the RPU. 
We show that our most efficient parameter choices can accelerate varying size (i)NTTs by up to $1485 \times$ over a traditional 32-core 2.5GHz AMD EPYC 7502 CPU. 

% \begin{figure}[h!]
% \begin{center}
% \centerline{\includegraphics[width=\columnwidth]{figures/haac_rpu_acc.pdf}}
% \vspace{-1em}
% \caption{SysOpt + HAAC + RPU - TinyImageNet ResNet-18}
% \label{fig:haac_rpu}
% \vspace{-1em}
% \end{center}
% \end{figure} 

Figure~\ref{fig:performance} (+RPU) shows that the HE evaluation latency is largely eliminated.
% We find that the HE evaluation latency is largely eliminated by RPU in Figure~\ref{fig:performance} (+RPU). 
Overall, by performing the system-level optimization, together with the hardware accelerators HAAC and RPU, we significantly reduced the end-to-end private inference latency by 76\% against the baseline.
The computation overhead is inconspicuous compared to the communication, which will be resolved by developing faster bandwidth and network technology in the future.

% shows that after accelerating GCs with HAAC, we further reduce the total latency for a private inference on ResNet18.
% The percentage of GCs in the computation latency is already very low, leaving the HE evaluation as a large occupation.
% the performance gain from the system-level optimization. 
% The overall latency is significantly reduced thanks to the optimized protocol.
% In the offline phase, the HE evaluation that occupied the majority of the latency in the baseline is greatly decreased, as well as the GCs evaluation for the online phase.
% However, besides communication, the computation for HE and GCs (i.e., the HE evaluation and GCs garbling) now dominate the total latency, which motivate us to look for a faster HE and GCs solution from other aspects.
% the total end-to-end latency of performing a single private inference on ResNet18 with a TinyImageNet input. 
% During the offline phase, the HE evaluations dominate the runtime while during the online phase, most compute cycles are spent on client-side GC evaluation. 

\section{RELATED WORKS} 
\label{sec:Related}
% \fixme{1 page}
% \fixme{Cambridge, Karthik, Austin, Negar}

% We now give a summary of seminal work in privacy-preserving computation technology and PI.

\textbf{PI:}
Prior work has explored using HE only, which is convenient as privacy primitives are not changed~\cite{Cryptonets, Craterlake}. 
However, these protocols cannot leverage LPHE and introduce accuracy loss via the approximation of ReLU, even with complex training~\cite{sisyphus}. 
% To overcome this, many have proposed hybrid protocols that improve upon these limitations. 
Additionally, some prior work includes a trusted-third party, which assumes a weaker security model for higher performance~\cite{Falcon,Cryptflow,SecureNN}.
The machine learning community has started exploring ways to design neural networks with fewer nonlinear function, such as pruning ReLUs from the networks~\cite{SNL, DeepReDuce}, and approximating ReLU computations for cheaper GCs implementations~\cite{Circa}.
DELPHI~\cite{delphi} and AESPA~\cite{AESPA} replace ReLUs with polynomial activation functions that are processed using Beaver Triples~\cite{BeaverTriples}, which are cheaper in both compute and communication than GCs. 
However, replacing ReLUs with low-degree polynomials reduces test accuracy, especially for deeper networks~\cite{sisyphus}. 
Therefore, we only considers highly accurate ReLU-only deep learning models that are state-of-the-art.

\textbf{GCs:}
% Prior work has also looked at accelerating GCs.
There are other prior works to accelerate GCs with GPU~\cite{GPU_GC, GPU_GC_malicious} and FPGA~\cite{Maxelerator, 2019FASE, GC_FPGA_Overlay, Garbledcpu}. 
We note that prior work uses the less secure fixed-key GCs setup~\cite{Half_Gate_rekey}, or uses SHA-1 instead of AES, which is simpler and less secure~\cite{GC_FPGA_Overlay}.
Moreover, most prior work uses small benchmarks that do not stress off-chip bandwidth, which is one of HAAC's primary contributions.
The uniqueness in HAAC when comparing against prior work is that HAAC considers parallel processing and pipelining at the same time, and also optimizes for off-chip communication. 
% Prior work does not consider parallel processing units and pipelining at the same time, and most do not optimize for off-chip communication. 
To the best of our knowledge, HAAC is the first ASIC GCs accelerator.
HAAC outperforms all prior accelerator and GPU works as shown in the paper~\cite{HAAC}.
% , such as Overlay~\cite{}, MAXelerator~\cite{Maxelerator}, FASE~\cite{2019FASE}. 
% The solutions include general accelerator~\cite{2019FASE, GC_FPGA_Overlay, Garbledcpu}, and application specific custom GCs hardware for MAC~\cite{Maxelerator}.
% Key insights are shown when comparing to the prior works.
% FASE orders gates by fanout hoping to resolve dependence and reduce stalls.
% HAAC's compiler performs analysis in data dependence graph, optimizing for ILP.
% Another real challenge of GCs computing emerges when considering off-chip effects.
% HAAC decouples gate execution and off-chip accesses, relieves the stress in memory traffic to support potential large workloads, while the prior works assume data fits on-chip. 
% These are where the major HAAC contributions shine.
% Overall, HAAC eliminates most of the performance overheads of GCs. 
% Additional compiler optimizations, higher levels
% of parallelism  and processing-in-memory (PIM) may help the future GCs implementation.

\textbf{HE:}
Several accelerators have been developed to improve the performance of HE primitives. F1 \cite{F1_2021} designs specialized functional units to accelerate primitive computations, such as NTT. However, F1 has a maximum polynomial degree support of only 16K, whereas RPU has no such limitations.
More recent HE accelerators such as CraterLake ~\cite{Craterlake}, BTS \cite{BTS}, and ARK ~\cite{ARK} target high multiplicative depth applications, but require large on-chip memories, 
% 256MB for Craterlake and 512MB for BTS and ARK. 
e.g., 256MB for Craterlake. 
Craterlake supports up to 64K polynomial degree. However, to support 128K ciphertext, the hardware needs modifications, which results in an additional area of 27.4$mm^2$, making it larger than RPU. Unlike Craterlake, RPU is flexible to support larger polynomials without hardware changes.

\section{DISCUSSION AND CONCLUSION}
\label{sec:Conclusion}

This paper presents solutions for improving privacy and security via PPCs.
The characterization identifies the need to completely redesign computing stacks from algorithms to storage, in order to enable practical private inference.
Point solutions are presented to demonstrate how well custom hardware can perform in overcoming many of the computational overheads leveraging classic architectural mechanisms: vectors (RPU) and VLIW (HAAC).
We conclude with a discussion of three predictions about what the future of PPC and systems for PPC will look like.

First, after rigorously studying private inference for multiple years, we feel hybrid protocols will be most valuable.
All PPC technologies (e.g., HE, GCs, SS, and OT) have strengths and critical weaknesses.
By effectively combining them and tailoring their use to workloads and threat models, one can often leverage their strengths while overcoming their weaknesses.
A prime example is ReLU and HE.
HE is a necessity for computing linear layers in ML, whether used directly online to process inputs or offline to compute secrets for SS.
However, popular integer/fixed-point schemes amenable for linear layer processing are not capable of processing non-linear functions.
Thus, HE can be combined with GCs to properly execute ReLU and preserve network accuracy.
We hope it inspires the community to consider accelerators beyond HE.

Second, through understanding the degree of slowdown at all aspects of a system--compute, communication, and storage--it is clear that the problem exceeds beyond custom hardware.
Hardware accelerators are certainly a necessity for overcoming the computational slowdowns in PPCs like GCs and HE, but are not alone sufficient.
To truly realize real-time private inference systems will need to leverage extremely dense storage technologies on the client device.
Communication between client and cloud will have to take place over the highest bandwidth wireless protocols offering the most bandwidth, e.g., up to 100 Gbps~\cite{ted2019wireless,ted20215G}.
Finally, even this may not be enough, and we need to re-think how we express problems as programs.
Looking again at PI, this can be seen in the need to re-think neural architectures to minimize ReLU counts, also known as the ReLU budget~\cite{CryptoNAS}.
Prior work has shown promising solutions that other application domains can learn from~\cite{Sphynx, SNL, CryptoNAS, DeepReDuce}.

Finally, we feel PPCs are the perfect application of co-design techniques.
In PPCs, all program behavior and information is known at compile time, i.e., programs are data oblivious.
This provides the compiler an ideal view of execution, and the opportunity to schedule computations and data movement just as well as any dynamic hardware.
This is likely important, and we feel PPC accelerators need to be programmable and software driven. 
The protocols will continue to evolve, and new applications will continually be ported. 
An ISA can solve these issues.
At the same time, the performance and efficiency problems of ISAs for general-purpose computing can largely be overcome.
By pushing all scheduling to the compiler (VLIW), leveraging highly decoupled pipelines (DAE), and classic parallel computing techniques (vectors), we believe the efficiency and performance of ASICs can be achieved without over-fitting hardware to a particular implementation.


% \fixme{ACKNOWLEDGMENTS: To Robert, for the bagels. (Huge Rob fan)}
\begin{acks}
% This work was supported in part by the Applications Driving Architectures (ADA) Research Center, a JUMP Center co-sponsored by SRC and DARPA. Additionally, this research was funded by the Defense Advanced Research Projects Agency (DARPA), under the Data Protection in Virtual Environments program, contract HR0011-21-9-0003.  
% Ebel was partially supported by the NY State Center for Advanced Technology in Telecommunications (CATT).
% The views, opinions, and/or findings expressed are those of the authors and do not necessarily reflect the views of the sponsors.
We acknowledge support from the ADA Research Center, SRC and DARPA. 
Additionally, this research was under the DPRIVE program, contract HR0011-21-9-0003.
The views expressed are those of the authors and do not necessarily reflect those of the sponsors.
\end{acks}

% \nobalance

\printbibliography

@inproceedings{Maxelerator,
  title={Maxelerator: FPGA accelerator for privacy preserving multiply-accumulate (MAC) on cloud servers},
  author={Hussain, Siam U and Rouhani, Bita Darvish and Ghasemzadeh, Mohammad and Koushanfar, Farinaz},
  booktitle={Proceedings of the 55th Annual Design Automation Conference},
  pages={1--6},
  year={2018}
}

@inproceedings{2019FASE,
  title={FASE: FPGA acceleration of secure function evaluation},
  author={Hussain, Siam U and Koushanfar, Farinaz},
  booktitle={2019 IEEE 27th Annual International Symposium on Field-Programmable Custom Computing Machines (FCCM)},
  pages={280--288},
  year={2019},
  organization={IEEE}
}

@article{HAAC,
%   title={HAAC: A Hardware-Software Co-Design to Accelerate Garbled Circuits},
%   author={Mo, Jianqiao and Gopinath, Jayanth and Reagen, Brandon},
%   journal={arXiv preprint arXiv:2211.13324},
%   year={2022}
% }

@inproceedings{VIP-Bench,
  title={VIP-Bench: A Benchmark Suite for Evaluating Privacy-Enhanced Computation Frameworks},
  author={Biernacki, Lauren and Demissie, Meron Zerihun and Workneh, Kidus Birkayehu and Namomsa, Galane Basha and Gebremedhin, Plato and Andargie, Fitsum Assamnew and Reagen, Brandon and Austin, Todd},
  booktitle={2021 International Symposium on Secure and Private Execution Environment Design (SEED)},
  pages={139--149},
  year={2021},
  organization={IEEE}
}

@inproceedings{Half_Gate_rekey,
  title={Better concrete security for half-gates garbling (in the multi-instance setting)},
  author={Guo, Chun and Katz, Jonathan and Wang, Xiao and Weng, Chenkai and Yu, Yu},
  booktitle={Advances in Cryptology--CRYPTO 2020: 40th Annual International Cryptology Conference, CRYPTO 2020, Santa Barbara, CA, USA, August 17--21, 2020, Proceedings, Part II},
  pages={793--822},
  year={2020},
  organization={Springer}
}

@article{FacebookTrillions,
  title={Accelerating facebook’s infrastructure with application-specific hardware},
  author={Lee, Kevin and Rao, Vijay and Arnold, William Christie},
  journal={Facebook. Retrieved August},
  volume={20},
  pages={2020},
  year={2019}
}

@inproceedings{delphi,
  title={Delphi: a cryptographic inference system for neural networks},
  author={Mishra, Pratyush and Lehmkuhl, Ryan and Srinivasan, Akshayaram and Zheng, Wenting and Popa, Raluca Ada},
  booktitle={Proceedings of the 2020 Workshop on Privacy-Preserving Machine Learning in Practice},
  pages={27--30},
  year={2020}
}

@article{GCintro1,
  title={A gentle introduction to yao’s garbled circuits},
  author={Yakoubov, Sophia},
  journal={preprint on webpage at https://web. mit. edu/sonka89/www/papers/2017ygc. pdf},
  year={2017}
}

@article{GCintro2,
  title={On Garbled Circuits},
  author={Navarro, Ignacio},
  year={2018}
}

@inproceedings{Half_Gate,
  title={Two halves make a whole: Reducing data transfer in garbled circuits using half gates},
  author={Zahur, Samee and Rosulek, Mike and Evans, David},
  booktitle={Advances in Cryptology-EUROCRYPT 2015: 34th Annual International Conference on the Theory and Applications of Cryptographic Techniques, Sofia, Bulgaria, April 26-30, 2015, Proceedings, Part II 34},
  pages={220--250},
  year={2015},
  organization={Springer}
}

@inproceedings{FreeXOR,
  title={Improved garbled circuit: Free XOR gates and applications},
  author={Kolesnikov, Vladimir and Schneider, Thomas},
  booktitle={Automata, Languages and Programming: 35th International Colloquium, ICALP 2008, Reykjavik, Iceland, July 7-11, 2008, Proceedings, Part II 35},
  pages={486--498},
  year={2008},
  organization={Springer}
}

@article{ObliviousTransfer,
  title={How to exchange secrets with oblivious transfer},
  author={Rabin, Michael O},
  journal={Cryptology ePrint Archive},
  year={2005}
}

@article{2020TFHE,
  title={TFHE: fast fully homomorphic encryption over the torus},
  author={Chillotti, Ilaria and Gama, Nicolas and Georgieva, Mariya and Izabach{\`e}ne, Malika},
  journal={Journal of Cryptology},
  volume={33},
  number={1},
  pages={34--91},
  year={2020},
  publisher={Springer}
}

@inproceedings{Yao1986GC,
  title={How to generate and exchange secrets},
  author={Yao, Andrew Chi-Chih},
  booktitle={27th annual symposium on foundations of computer science (Sfcs 1986)},
  pages={162--167},
  year={1986},
  organization={IEEE}
}

@article{Sphynx,
  title={Sphynx: A Deep Neural Network Design for Private Inference},
  author={Cho, Minsu and Ghodsi, Zahra and Reagen, Brandon and Garg, Siddharth and Hegde, Chinmay},
  journal={IEEE Security \& Privacy},
  volume={20},
  number={5},
  pages={22--34},
  year={2022},
  publisher={IEEE}
}

@article{Stochastic_TFHE,
  title={Homomorphically Encrypted Computation using Stochastic Encodings},
  author={Hsiao, Hsuan and Lee, Vincent and Reagen, Brandon and Alaghi, Armin},
  journal={arXiv preprint arXiv:2203.02547},
  year={2022}
}

@inproceedings{GC_FPGA_Overlay,
  title={Secure function evaluation using an fpga overlay architecture},
  author={Fang, Xin and Ioannidis, Stratis and Leeser, Miriam},
  booktitle={Proceedings of the 2017 ACM/SIGDA International Symposium on Field-Programmable Gate Arrays},
  pages={257--266},
  year={2017}
}

@article{ted2019wireless,
  title={Wireless communications and applications above 100 GHz: Opportunities and challenges for 6G and beyond},
  author={Rappaport, Theodore Scott and Xing, Yunchou and Kanhere, Ojas and Ju, Shihao and Madanayake, Arjuna and Mandal, Soumyajit and Alkhateeb, Ahmed and Trichopoulos, Georgios C},
  journal={IEEE access},
  volume={7},
  pages={78729--78757},
  year={2019},
  publisher={IEEE}
}

@misc{swanky_fancyGC,
  title={swanky: A suite of rust libraries for secure multi-party computation},
  author={Carmer, Brent and Malozemoff, Alex J and Rosen, Marc},
  year={2019}
}

@article{ted20215G,
  title={5G's Killer App Will Be 6G: Massive MIMO Millimeter Waves and Small Cell Infrastructure Will Pay Off for Future Tech Generations},
  author={Rappaport, TS},
  journal={IEEE Spectrum OP-ED},
  year={2021}
}

@inproceedings{Cryptflow,
  title={Cryptflow: Secure tensorflow inference},
  author={Kumar, Nishant and Rathee, Mayank and Chandran, Nishanth and Gupta, Divya and Rastogi, Aseem and Sharma, Rahul},
  booktitle={2020 IEEE Symposium on Security and Privacy (SP)},
  pages={336--353},
  year={2020},
  organization={IEEE}
}

@InProceedings{BeaverTriples,
author="Beaver, Donald",
editor="Coppersmith, Don",
title="Precomputing Oblivious Transfer",
booktitle="Advances in Cryptology --- CRYPT0' 95",
year="1995",
publisher="Springer Berlin Heidelberg",
address="Berlin, Heidelberg",
pages="97--109",
isbn="978-3-540-44750-4"
}

@article{AESPA,
  title={AESPA: Accuracy preserving low-degree polynomial activation for fast private inference},
  author={Park, Jaiyoung and Kim, Michael Jaemin and Jung, Wonkyung and Ahn, Jung Ho},
  journal={arXiv preprint arXiv:2201.06699},
  year={2022}
}

@article{Circa,
  title={Circa: Stochastic relus for private deep learning},
  author={Ghodsi, Zahra and Jha, Nandan Kumar and Reagen, Brandon and Garg, Siddharth},
  journal={Advances in Neural Information Processing Systems},
  volume={34},
  pages={2241--2252},
  year={2021}
}

@article{Falcon,
author = {Wagh, Sameer and Tople, Shruti and Benhamouda, Fabrice and Kushilevitz, Eyal and Mittal, Prateek and Rabin, Tal},
year = {2021},
month = {01},
pages = {188-208},
title = {Falcon: Honest-Majority Maliciously Secure Framework for Private Deep Learning},
volume = {2021},
journal = {Proceedings on Privacy Enhancing Technologies},
doi = {10.2478/popets-2021-0011}
}

@article{SecureNN,
  title={SecureNN: 3-Party Secure Computation for Neural Network Training.},
  author={Wagh, Sameer and Gupta, Divya and Chandran, Nishanth},
  journal={Proc. Priv. Enhancing Technol.},
  volume={2019},
  number={3},
  pages={26--49},
  year={2019}
}

@inproceedings{Garbledcpu,
  title={Garbledcpu: a mips processor for secure computation in hardware},
  author={Songhori, Ebrahim M and Zeitouni, Shaza and Dessouky, Ghada and Schneider, Thomas and Sadeghi, Ahmad-Reza and Koushanfar, Farinaz},
  booktitle={Proceedings of the 53rd Annual Design Automation Conference},
  pages={1--6},
  year={2016}
}

@inproceedings{GPU_GC,
  title={GPU and CPU parallelization of honest-but-curious secure two-party computation},
  author={Husted, Nathaniel and Myers, Steven and Shelat, Abhi and Grubbs, Paul},
  booktitle={Proceedings of the 29th Annual Computer Security Applications Conference},
  pages={169--178},
  year={2013}
}

@inproceedings{GPU_GC_malicious,
  title={Faster maliciously secure two-party computation using the GPU},
  author={Frederiksen, Tore Kasper and Jakobsen, Thomas P and Nielsen, Jesper Buus},
  booktitle={Security and Cryptography for Networks: 9th International Conference, SCN 2014, Amalfi, Italy, September 3-5, 2014. Proceedings 9},
  pages={358--379},
  year={2014},
  organization={Springer}
}

@inproceedings{FHEW_like,
  title={Bootstrapping in FHEW-like cryptosystems},
  author={Micciancio, Daniele and Polyakov, Yuriy},
  booktitle={Proceedings of the 9th on Workshop on Encrypted Computing \& Applied Homomorphic Cryptography},
  pages={17--28},
  year={2021}
}

@inproceedings{SNL,
  title={Selective network linearization for efficient private inference},
  author={Cho, Minsu and Joshi, Ameya and Reagen, Brandon and Garg, Siddharth and Hegde, Chinmay},
  booktitle={International Conference on Machine Learning},
  pages={3947--3961},
  year={2022},
  organization={PMLR}
}

@inproceedings{GAZELLE,
  title={$\{$GAZELLE$\}$: A low latency framework for secure neural network inference},
  author={Juvekar, Chiraag and Vaikuntanathan, Vinod and Chandrakasan, Anantha},
  booktitle={27th $\{$USENIX$\}$ Security Symposium ($\{$USENIX$\}$ Security 18)},
  pages={1651--1669},
  year={2018}
}

@inproceedings{DeepReDuce,
  title={DeepReDuce: Relu reduction for fast private inference},
  author={Jha, Nandan Kumar and Ghodsi, Zahra and Garg, Siddharth and Reagen, Brandon},
  booktitle={International Conference on Machine Learning},
  pages={4839--4849},
  year={2021},
  organization={PMLR}
}

@inproceedings{Cheetah,
  title={Cheetah: Optimizing and accelerating homomorphic encryption for private inference},
  author={Reagen, Brandon and Choi, Woo-Seok and Ko, Yeongil and Lee, Vincent T and Lee, Hsien-Hsin S and Wei, Gu-Yeon and Brooks, David},
  booktitle={2021 IEEE International Symposium on High-Performance Computer Architecture (HPCA)},
  pages={26--39},
  year={2021},
  organization={IEEE}
}

@inproceedings{RPU,
%   title={RPU: The Ring Processing Unit},
%   author={Soni, Deepraj  and Neda, Negar and Zhang, Naifeng and Reynwa, Benedict  and Heyman, Benjamin  and Moopan, Mohammed Nabeel Thari  and Badawi, Ahmad Al  and Polyakov, Yuriy  and Canida, Kellie  and Schmidt, Andrew  and Pedram, Massoud  and Cousins, David Bruce  and French, Matthew and Franchetti, Franz and Maniatakos, Michail  and Reagen, Brandon },
%   booktitle={2023 IEEE International Symposium on Performance Analysis of Systems and Software (ISPASS)},
%   % pages={26--39},
%   year={2023},
%   organization={IEEE}
% }

@article{sisyphus,
  title={Sisyphus: A cautionary tale of using low-degree polynomial activations in privacy-preserving deep learning},
  author={Garimella, Karthik and Jha, Nandan Kumar and Reagen, Brandon},
  journal={arXiv preprint arXiv:2107.12342},
  year={2021}
}

@inproceedings{F1_2021,
  title={F1: A fast and programmable accelerator for fully homomorphic encryption},
  author={Samardzic, Nikola and Feldmann, Axel and Krastev, Aleksandar and Devadas, Srinivas and Dreslinski, Ronald and Peikert, Christopher and Sanchez, Daniel},
  booktitle={MICRO-54: 54th Annual IEEE/ACM International Symposium on Microarchitecture},
  pages={238--252},
  year={2021}
}

@inproceedings{SAFENet,
  title={SAFENet: A secure, accurate and fast neural network inference},
  author={Lou, Qian and Shen, Yilin and Jin, Hongxia and Jiang, Lei},
  booktitle={International Conference on Learning Representations},
  year={2021}
}

@article{CryptoNAS,
  title={Cryptonas: Private inference on a relu budget},
  author={Ghodsi, Zahra and Veldanda, Akshaj Kumar and Reagen, Brandon and Garg, Siddharth},
  journal={Advances in Neural Information Processing Systems},
  volume={33},
  pages={16961--16971},
  year={2020}
}

@inproceedings{Cryptonets,
  title={Cryptonets: Applying neural networks to encrypted data with high throughput and accuracy},
  author={Gilad-Bachrach, Ran and Dowlin, Nathan and Laine, Kim and Lauter, Kristin and Naehrig, Michael and Wernsing, John},
  booktitle={International conference on machine learning},
  pages={201--210},
  year={2016},
  organization={PMLR}
}

@inproceedings{Craterlake,
  title={Craterlake: a hardware accelerator for efficient unbounded computation on encrypted data},
  author={Samardzic, Nikola and Feldmann, Axel and Krastev, Aleksandar and Manohar, Nathan and Genise, Nicholas and Devadas, Srinivas and Eldefrawy, Karim and Peikert, Chris and Sanchez, Daniel},
  booktitle={Proceedings of the 49th Annual International Symposium on Computer Architecture},
  pages={173--187},
  year={2022}
}

@article{Spiral,
  author={Franchetti, Franz and Low, Tze Meng and Popovici, Doru Thom and Veras, Richard M. and Spampinato, Daniele G. and Johnson, Jeremy R. and Püschel, Markus and Hoe, James C. and Moura, José M. F.},
  journal={Proceedings of the IEEE}, 
  title={SPIRAL: Extreme Performance Portability}, 
  year={2018},
  volume={106},
  number={11},
  pages={1935-1968},
  doi={10.1109/JPROC.2018.2873289}}

@inproceedings{BTS,
author = {Kim, Sangpyo and Kim, Jongmin and Kim, Michael Jaemin and Jung, Wonkyung and Kim, John and Rhu, Minsoo and Ahn, Jung Ho},
title = {BTS: An Accelerator for Bootstrappable Fully Homomorphic Encryption},
year = {2022},
isbn = {9781450386104},
publisher = {Association for Computing Machinery},
address = {New York, NY, USA},
url = {https://doi.org/10.1145/3470496.3527415},
doi = {10.1145/3470496.3527415},
pages = {711–725},
numpages = {15},
location = {New York, New York},
series = {ISCA '22}
}

@inproceedings{ARK,
  title={Ark: Fully homomorphic encryption accelerator with runtime data generation and inter-operation key reuse},
  author={Kim, Jongmin and Lee, Gwangho and Kim, Sangpyo and Sohn, Gina and Rhu, Minsoo and Kim, John and Ahn, Jung Ho},
  booktitle={2022 55th IEEE/ACM International Symposium on Microarchitecture (MICRO)},
  pages={1237--1254},
  year={2022},
  organization={IEEE}
}

@inproceedings{hybrid_PI,
author={Garimella, Karthik and Ghodsi, Zahra and Jha, Nandan Kumar and Garg, Siddharth and Reagen, Brandon},
title={Characterizing and Optimizing End-to-End Systems for Private Inference},
year={2023},
%publisher={Association for Computing Machinery},
address={New York, NY, USA},
url={https://doi.org/10.1145/3582016.3582065},
doi={10.1145/3582016.3582065},
numpages={16},
location={New York, New York},
series={ASPLOS ’23}
}

@String{Computing = "Computing" }

@String{Computer = "{IEEE} Computer" }

@String{Springer = "Springer-Verlag" }


\end{document}